\documentclass[preprint]{vgtc}               % preprint
\graphicspath{{figures/}{pictures/}{images/}{./}} % where to search for the images

\usepackage{times}                     % we use Times as the main font
\usepackage{tabu}                      % only used for the table example
\usepackage{booktabs}                  % only used for the table example
\usepackage{lipsum}                    % used to generate placeholder text
\usepackage{mwe}                       % used to generate placeholder figures

\usepackage{mathptmx}                  % use matching math font
\usepackage{balance}

\newcommand{\rev}[1]{{#1}}

\onlineid{0}

\vgtccategory{Research}

\vgtcinsertpkg

\ieeedoi{10.1109/VIS60296.2025.00043}

\title{Data-Driven Compute Overlays for Interactive \\Geographic Simulation and Visualization}

\author{Patrick Komon\thanks{e-mail: pkomon@cg.tuwien.ac.at}\\ %
        \scriptsize TU Wien %
\and Gerald Kimmersdorfer\thanks{e-mail: gkimmersdorfer@cg.tuwien.ac.at}\\ %
     \scriptsize TU Wien %
\and Adam Celarek\thanks{e-mail: celarek@cg.tuwien.ac.at}\\ %
     \scriptsize TU Wien %
\and Manuela Waldner\thanks{e-mail: waldner@cg.tuwien.ac.at}\\ %
     \scriptsize TU Wien}

\teaser{
  \centering
  \includegraphics[alt={Mountains in Tyrol. Three areas near the peak of a mountain in the foreground are highlighted in blue. Originating from these areas, a flow representing the avalanche is heading downslope along the surface. The avalanche reaches speeds of up to 70 meters per second. Upon reaching the valley, it slows down and stops after a short distance.},width=\linewidth]{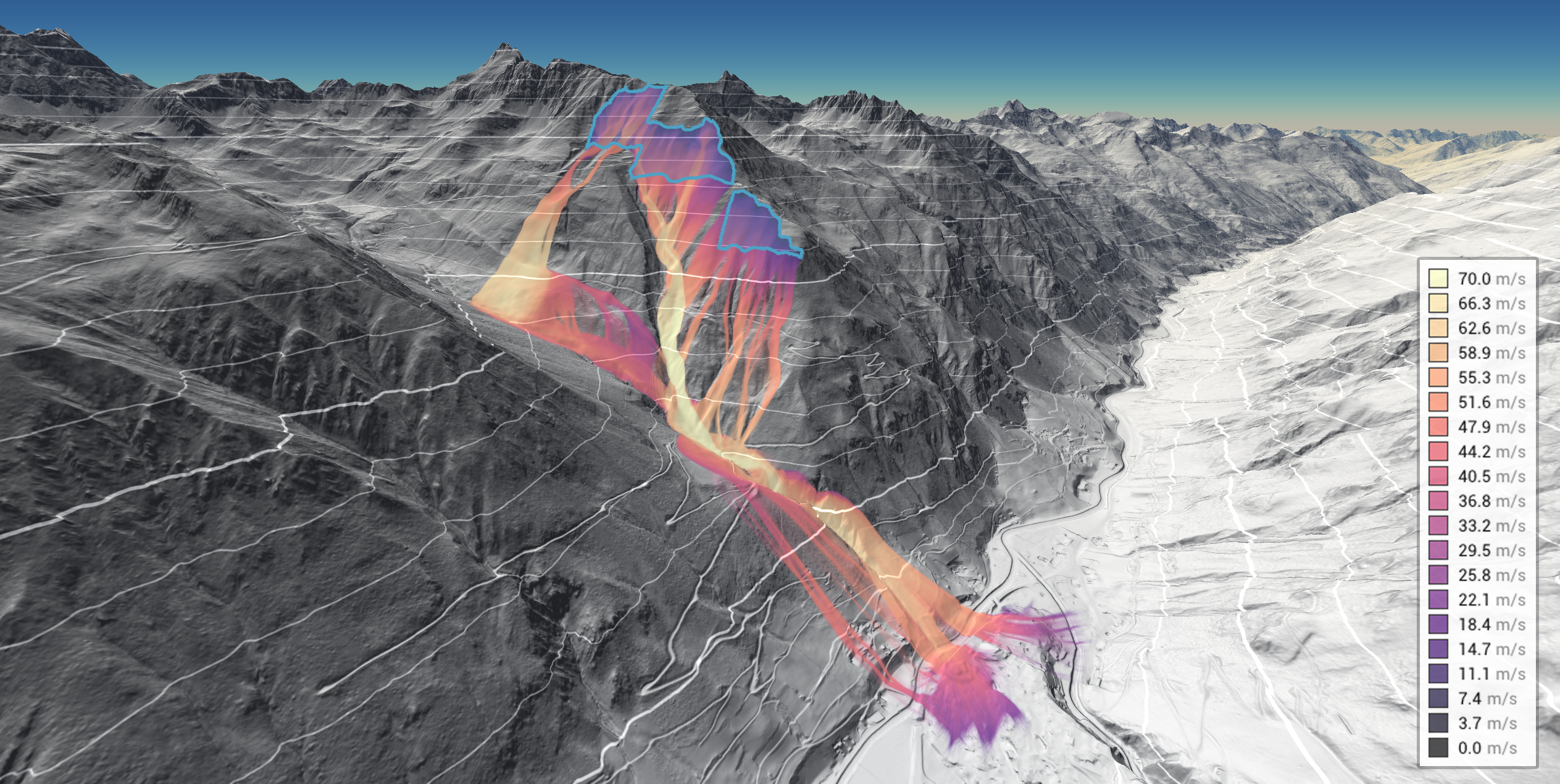}
  \caption{\rev{{Simulation of the Madlein avalanche (Tyrol, 1984) from its recorded release areas (highlighted in blue), computed and visualized in a few milliseconds. The color encodes the velocity.} }}
  \label{fig:teaser}
}

\abstract{
    We present interactive data-driven compute overlays for native and web-based 3D geographic map applications based on WebGPU. Our data-driven overlays are generated in a multi-step compute workflow from multiple data sources on the GPU. We demonstrate their potential by showing results from snow cover and avalanche simulations, where simulation parameters can be adjusted interactively and results are visualized instantly. Benchmarks show that our approach can compute large-scale avalanche simulations in milliseconds to seconds, depending on the size of the terrain and the simulation parameters, which is multiple orders of magnitude faster than a state-of-the-art Python implementation. 
} % end of abstract


\begin{document}


\firstsection{Introduction}

\maketitle

Data-driven overlays for geographic maps showing, for instance, steepness of the terrain, precipitation, or points of interest, are typically computed offline. This limits the applicability to static data. To overlay 2D or 3D maps with visualizations of dynamic data or results of interactively steerable simulations, the overlays either need to be limited to a very small area or require remote access to a powerful server.

In this work, we explore the use of WebGPU to compute data-driven overlays \rev{for online maps} dynamically with the support of compute shaders. We present a workflow for running simulations and instantly generating overlays for 3D maps and demonstrate how our prototype \emph{weBIGeo} can speed up conventional workflows by orders of magnitude. In summary, our contributions are as follows:  
\begin{enumerate}
    \item \rev{An API to specify} compute workflow\rev{s} for instant generation of data-driven \rev{multi-resolution} overlays for 3D maps. 
    \item The first open-source 3D geographic simulation and visualization tool \emph{weBIGeo} for interactive native and web-based applications with compute shader support, 
    \item Two use cases -- snow cover and avalanche simulation (\cref{fig:teaser}) -- demonstrating the potential of data-driven overlays. 
\end{enumerate}
Our source code is available on \rev{\href{https://github.com/weBIGeo/webigeo}{GitHub}}. An online demo showcasing the avalanche simulation use case can be accessed here: \url{https://webigeo.alpinemaps.org/} (WebGPU-capable browser required, for example Google Chrome \rev{on Windows}).

\section{Background: (Web-Based) GeoVis}

% \begin{figure*}[ht]
%     \centering
%     \includegraphics[width=1\linewidth]{figures/general_compute_graph.pdf}
%     \caption{An exemplary compute workflow, which consists of two compute steps and three distinct inputs, each with its own input type. An optional LOD generation step produces a multi-resolution output texture.}
%     \label{fig:compute-pipeline}
% \end{figure*}

\begin{figure*}[ht]
    \centering
    \includegraphics[alt={A node graph specifying the avalanche workflow. Nodes have names, inputs and outputs. The outputs of each node are connected to inputs of subsequent nodes.},width=1\linewidth]{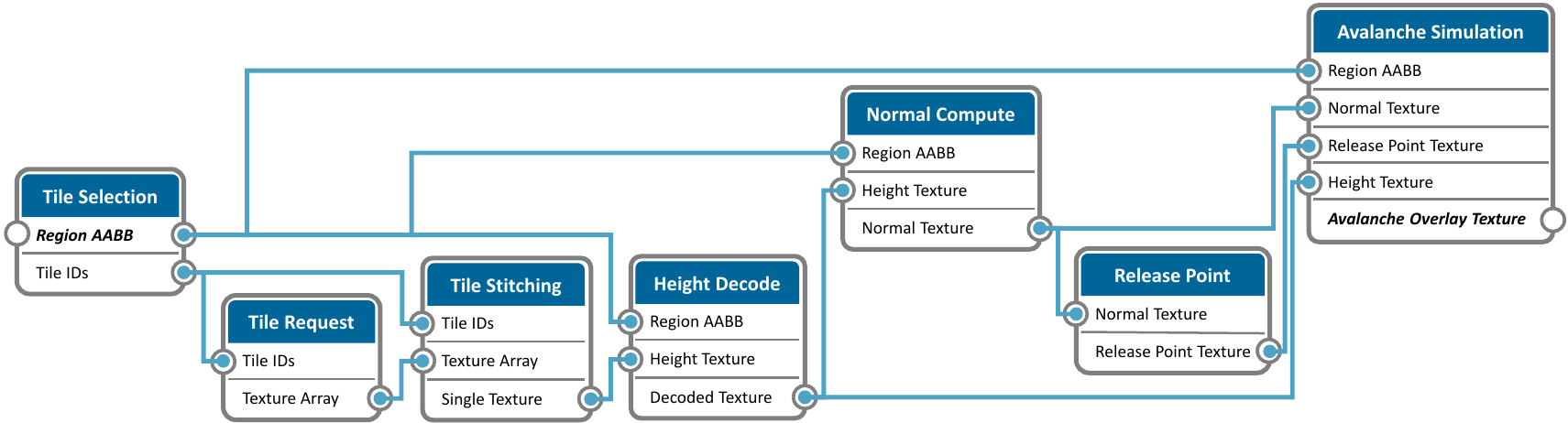}
    \caption{\rev{{A node graph generated from the workflow specification for the avalanche use case (\cref{sec:avalanches}) This workflow generates a multi-resolution \emph{Avalanche Overlay Texture} (right) from terrain model tiles in a given \emph{Region AABB} (left).}}}
    \label{fig:compute-pipeline}
\end{figure*}

%There are multiple ways how to provide interactive geographic visualizations. 
Common geographic visualization techniques are glyphs placed on top of maps and color overlays~\cite{luo2003visualizing}. The focus of our work lies on overlays. Here, we review three principle paradigms to data-driven map overlays particularly for the web: 

%The most straight-forward method is to \rev{provide data-driven overlays as pre-rendered images} on a dedicated server. Overlays are served as \rev{tiles, i.e., a level-of-detail structure of rasterized images}. Conceptually, such 
\rev{Map tiles} are the foundation for all popular online map applications, where map clients request pre-rasterized map images depending on the currently shown zoom level and visible field of view~\cite{quinn2010predictive} \rev{from a dedicated server} -- also known as ``slippy maps''~\cite{haklay2008openstreetmap}. Data-driven overlays can be similarly \rev{pre-}rendered into \rev{\textbf{ offline tiles}} and overlaid onto a base map\rev{~\cite{egenhofer1993exploratory,reddy1999terravision}}. A well-known open-source library to build layered maps is OpenLayers~\cite{santiago2015book}. Offline tiles are useful when the data to be overlaid itself is big and static. %A simple client application can request tiles of an appropriate quality and thereby control the amount of data to be downloaded from the tile server. 
The disadvantage \rev{is} that the server has to re-compute the tiles whenever the underlying data changes. Users have no or only very limited possibilities to interact with the \rev{data.} %, for instance by filtering or changing simulation parameters. 
%Commercial platforms compute overlays in powerful data centers to provide faster responses. However, it is not the ideal solution if the user wants to interact with the data itself, for instance perform filtering or adjusting the parameters of a simulation. 

An alternative approach is to render the final images to be viewed by the client directly on the server. The client only forwards users' interaction with the map (e.g., panning and zooming) and with the data (e.g., filtering) to the server and finally shows the image transmitted by the server. The application is therefore largely independent from the client's hardware resources. %The amount of data to be transferred depends on the user interaction and the screen size and is independent of the size of the data to be analyzed. 
In the geographic context, examples for such a \textbf{server-side analyis and rendering} approach are HORA 3D for interactive flooding simulations~\cite{rauer2024hora} and applications built on top of ParaView~\cite{cedilnik2006remote}. The disadvantage is that the server needs to be powerful enough to serve all requests at the same time. In addition, interactive applications require a lot of bandwidth for smooth interaction, such as flying over the 3D terrain in \cref{fig:teaser}. 

The most flexible approach is \textbf{client-side analysis and rendering}: the data to be analyzed remains on the client, and the client handles all interactivity, simulation, and rendering itself. The major disadvantage, clearly, is that the client needs to be powerful enough -- in terms of CPU, GPU, and memory -- to handle the computation of the overlays and all user interactivity. 
Client-side overlays can be computed pixel-wise for every frame in the fragment shader based on the raster tiles used for rendering the 3D map. Examples are \rev{risk maps computed from local terrain properties~\cite{eschner-2023-evl} or difference maps of flood simulation scenarios~\cite{heitzler2017gpu}.}
%local steepness overlays or an avalanche risk map based on the local steepness, exposition, height, and the local avalanche danger level. 
There are \rev{some} major shortcomings to this approach: First, overlays need to be recomputed whenever user navigation requires a new map tile to be loaded, even if the underlying data does not change. Second, the resolution of the overlay is bound to the resolution of the underlying map tile. \rev{Third, pixel-wise analysis is not sufficient for more complex scenarios like avalanche simulation (see~\cref{sec:avalanches}).}
%The map tile resolution depends on the camera distance and is therefore not uniform across the screen for 3D maps -- in contrast to 2D maps. Computing overlays in screen resolution may lead to artifacts and information loss, as we will illustrate in \cref{sec:snow}. 
%It is therefore desirable that client-side overlays are computed only on demand (i.e., if the underlying data changes) and in an appropriate, homogeneous resolution independent of the currently displayed map resolution. 
\rev{It has been shown that some more complex geospatial analyses can be significantly sped up by GPGPU, using, for instance OpenCL~\cite{dubel2017visualizing} or CUDA~\cite{xia2011accelerating}. Vollmer et al.~\cite{vollmer2018hierarchical} compute levels-of-detail of pre-computed static overlay textures on the GPU so that the user can interactively change aggregation parameters. In contrast, we propose a workflow and present an implementation to interactively compute overlay textures directly from the raw data. }

%An example for map-independent client-side overlays is presented by  PrioVis~\cite{dubel2017visualizing}, which partially utilizes OpenCL for parallel data analysis on the GPU. 
%However, up 
\rev{Up} until recently, graphics APIs for the web were limited to vertex and fragment shaders. With the new web-based graphics and compute GPU API \emph{WebGPU}, it is now possible to employ compute shaders in the browser. 
%With \emph{weBIGeo}, we explore the potential and limitations of client-side WebGPU-based data-driven overlays for 3D maps.
Web\-GPU is increasingly generating interest for interactive web-based visualization of large data. In particular, it has been used for interactive isosurface extraction~\cite{usher2020interactive}, graph drawing~\cite{dyken2022graphwagu}, and aggregate 2D GeoVis~\cite{kimmersdorfer2023webgpu}. Web\-GPU is generally seen as a performant successor of WebGL in the context of Web\-GIS applications~\cite{usta2024webgpu} \rev{and is appreciated for its code portability between web, desktop, and mobile map applications~\cite{ammann2022maplibre}}. We extend these works by contributing a workflow, a software architecture, and performance benchmarks for data-driven compute overlays for 3D maps. \rev{To the best of our knowledge, \emph{weBIGeo} is thereby the first online map with interactively computed data-driven overlays. }

\section{Data-Driven Compute Overlays}
\label{sec:overlays}

\rev{\emph{weBIGeo} exposes a simple and flexible API to construct compute workflows of varying complexity. Compute workflows are specified as a sequence of compute nodes, similarly to dataflow models for visualization provenance~\cite{silva2007provenance} or for steering simulation ensembles~\cite{waser2011nodes}. Each compute node has at least one input and one output, as illustrated in the node graph in \cref{fig:compute-pipeline}.} Raw input data can come in various forms -- from single values, such as the current snow line in the region of interest, over spatial point data stored in a spatial data structure, such as the location of weather stations stored in a kd-tree, to raster data stored in a tile format, such as a digital elevation model \rev{(DEM)} capturing the ground elevation \rev{within a given region}. The output of a compute step can be \rev{resources like} textures, buffers, or vectors. The compute workflow can be configured such that the output of one compute step can serve as input to a subsequent one. This allows arbitrarily reusing and composing compute steps across multiple compute workflow definitions. 
%\rev{Compute nodes themselves can represent any CPU or GPU computation, if the data can be decomposed} small, independent units \rev{that can be executed in} a compute shader for each unit in parallel. 
\rev{Compute nodes themselves can represent any CPU or GPU computation. If the input data can be decomposed into small, independent units, a node may execute a compute shader to process each unit in parallel.}

\emph{weBIGeo} employs a multi-pass rendering pipeline with deferred shading. 
In the geometry pass, it draws the z-buffer of the base map with albedo, position, and normal. 
In the compute pass, the multi-resolution overlay texture is generated, \rev{by executing the specified compute workflow}. 
In the final compose pass, we first set an atmosphere background, then apply Phong shading to the base map, and finally \rev{alpha-}blend the data-driven overlay with the base map \rev{at its specified location}. 
%The blending is currently implemented as simple alpha blending. 
% May be limiting; base map may interfere with thematic overlay~\cite{kunz2011enhance}

\rev{A major advantage of compute overlays in} contrast to an approach that computes overlays in \rev{the geometry pass using a} fragment shader \rev{ is that the compute pass only needs to be re-executed} when the input data changes or the user has modified simulation parameters.
\rev{Another advantage is that data-driven overlays can be computed} on the finest reasonable zoom level \rev{of the input data}, independently of the base map resolution\rev{, which }depends on the camera distance and is therefore not uniform across the screen for 3D maps. Computing overlays in \rev{map} resolution may lead to artifacts and information loss, as we will illustrate in \cref{sec:snow}. %\rev{In addition, the user can control the output resolution of the generated overlay texture.}  
From the high-resolution \rev{overlay} texture, we then compute a mipmap\rev{, which} represents the data-driven, multi-resolution overlay.

\rev{A disadvantage of compute overlays is that the entire high-resolution overlay texture needs to be computed in a single compute pass. 
This implies that} the extent of the region for which to compute the overlay is bound by the available client resources, most importantly GPU memory and maximum supported texture size. In our implementation, we \rev{therefore} limit the texture size to WebGPU's current default (minimum) limit of $8192 \times 8192$ pixels to ensure compatibility with all devices supporting WebGPU. 

%\PK{An example would be nice here. For example: snow in the highest resolution: how large can the region be for one of your benchmark computers? How much memory do they have / do we need?}
%Another disadvantage is that overlays are only stored on the GPU by default. Reading back the overlay textures to the CPU for further use is possible, but it causes additional computational overhead. \PK{some rough numbers here?}

%\TODO{do you want to discuss how this flexibility enables us to have a visual node editor for compute shader configuration? With a screenshot maybe even?}

%Representing computation steps of the compute workflow as nodes in a directed-acyclic graph, as illustrated in \cref{fig:compute-pipeline}, has multiple advantages. The interface of each compute step is define by the set of inputs and outputs, each with a specific data type, that can be connected to other inputs ad outputs of other compute steps. This allows arbitrarily reusing and composing compute steps across multiple compute workflow definitions. Additionally, it enables compute workflows to be visualized and configured in a visual node editor in an easy-to-understand way.

\section{\rev{Implementation}}
\label{sec:weBIGeo}

\rev{Conceptually, the data-driven compute overlay API of \emph{weBIGeo} could be implemented on top of any graphics API or rendering engine that supports compute shaders. With WebGPU, we now for the first time have the opportunity to provide client-side compute overlays for web-based map applications. }

%\emph{weBIGeo} uses a hybrid approach, utilizing both, pre-computed tiles and on-the-fly client-side analysis and rendering. 
For rendering the 3D base map, \emph{weBIGeo} uses a traditional tile-based map rendering approach. As input, it requests tiles from a \rev{DEM} to compute the 3D terrain mesh, as well as corresponding orthophotos as textures. At the moment, \emph{weBIGeo} is limited to Austria and uses open data provided by \href{https://basemap.at/}{basemap.at}. 

\emph{weBIGeo} is written in C++ on top of \rev{our} 3D map rendering framework \href{https://github.com/AlpineMapsOrg}{AlpineMaps.org} that takes care of camera handling, tile scheduling, and caching. We additionally use \href{https://www.qt.io/}{Qt} for handling multi-threading, networking, file I/O etc., and \href{https://github.com/ocornut/imgui}{ImGUI} for the user interface. 
\emph{weBIGeo} can be built as a native Windows or Linux application (with Google's \href{https://dawn.googlesource.com/dawn}{Dawn} backend), or as a web application (using the \href{https://emscripten.org/}{Emscripten} compiler to convert the C++ code into \href{https://webassembly.org/}{Web\-Assembly} format, which is supported by most of today's browsers).
%\emph{weBIGeo} can be built as a native Windows application or as a web application. For native applications and for running the application in the Chrome browser, it uses Google's \href{https://dawn.googlesource.com/dawn}{Dawn} to execute Web\-GPU. To build a web application, we employ the \href{https://emscripten.org/}{Emscripten} compiler to convert the C++ code into \href{https://webassembly.org/}{WebAssembly} format, which is supported by most of today's browsers. 

% \emph{weBIGeo} employs a multi-pass rendering pipeline with deferred shading. 
% In the geometry pass, it draws the z-buffer of the base map with albedo, position, and normal. 
% In the compute pass, the multi-resolution overlay texture is generated, as described in \cref{sec:overlays} and illustrated in \cref{fig:compute-pipeline}. 
% In the final compose pass, we first set an atmosphere background, then apply Phong shading to the base map, and finally blend the data-driven overlay with the base map. The blending is currently implemented as simple alpha blending. 
% % May be limiting; base map may interfere with thematic overlay~\cite{kunz2011enhance}

\section{Use Cases}
\label{sec:useCases}

We \rev{show} two proof-of-concept \rev{compute workflows} to demonstrate the potential \rev{and flexibility} of data-driven compute overlays. \rev{For both use cases, we briefly discuss the compute workflow nodes, the simulation parameters, the data decomposition strategy, and the results.}
For the first example, we draw a visual comparison between an overlay computed in the fragment shader in base map resolution and our data-driven compute overlays. The second example shows the results of avalanche simulations, which cannot be computed in a conventional fragment shader approach. We report performance benchmarks for this use case and compare the results to a state-of-the-art avalanche simulation without GPU support. 

\subsection{Snow Cover}
\label{sec:snow}

\rev{The first use case is a} simple snow cover simulation, which predicts snow cover solely from terrain features and thereby can easily achieve a fast visual effect for 3D maps. 
Through the GUI, the user can interactively control a few parameters, such as the altitude of the snow line, the maximum steepness of the terrain, as well as blending factors for the transition areas. The snow cover overlay renders snow-covered areas in white, while the alpha corresponds to the blending factor. 

\rev{The first five steps of the compute workflow are identical to the workflow for simulating avalanches, discussed in \cref{sec:avalanches} and shown in \cref{fig:compute-pipeline}: Within a desired region bounding box, we retrieve corresponding high-resolution} (approximately $1 \times 1$ meters) \rev{tiles of the DEM, stitch the tiles into one DEM texture, and compute the surface normals. The final compute node then decides }-- depending on the given input parameters -- whether to set the albedo of the base map to white or whether to pick the color of the corresponding orthophoto texture. \rev{The normals and final snow computation can both be easily parallelized for each texel of the DEM texture.} Given \rev{WebGPU's texture size} limitations, we can currently cover a region of $2048 \times 2048$ meters ($\approx 4.2km^2$) requiring 256 MiB of GPU memory. 

\begin{figure}
    \centering
    \includegraphics[alt={3D Terrain from above. A rectangular region is covered in snow. Top left: Detailed, no snow on fine terrain features such as trees. Top right: zoomed-out version of left, lower displayed map resolution because the terrain is farther away, snow covers entire area. Bottom left: Detailed. Bottom right: Snow overlay still detailed, despite lower displayed map resolution.},width=1\linewidth]{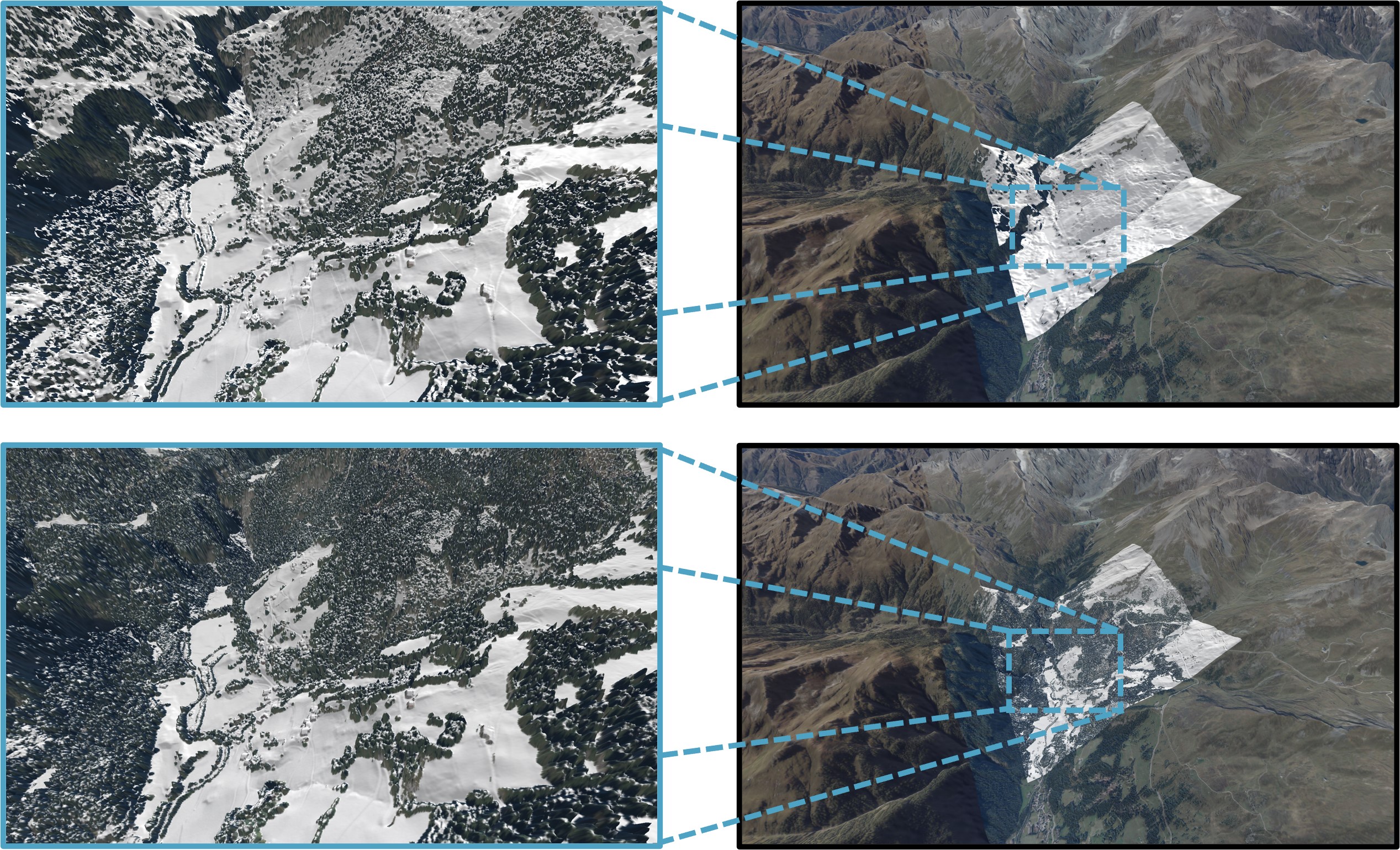}
    \caption{Simple snow cover simulation using a fragment shader on top of the base map (top) or using a data-driven compute overlay (bottom). When zooming out (right), the fragment shader approach shows too much snow on top of forests and other locally steep terrain (top right). \rev{Zoom in for more details. }}
    \label{fig:snow}
\end{figure}

Since the terrain is only evaluated locally \rev{for this snow effect, it could theoretically also be} computed in a fragment shader in each base map rendering step based on the \rev{DEM} geometry. %: The shader program decides -- depending on the given input parameters -- whether to set the albedo of the base map to white or whether to pick the color of the corresponding orthophoto texture. 
\cref{fig:snow} (top) illustrates the limitations of this approach: by coupling the snow cover simulation to the currently rendered base map resolution, the visual quality derogates significantly as the user zooms out. This happens because the normal vectors, based on which the steepness is evaluated, are flattened \rev{in map resolution as the distance to the camera increases, and more coarse DEM tiles are loaded}. Thus, regions with large variety in steepness from the \rev{DEM}, such as forests, will be interpreted as flat and rendered in white \rev{from distance}. In contrast, the data-driven snow cover overlay \rev{is computed on the highest available resolution of the DEM.} In the compose pass, the renderer then interpolates the overlay texture rather than the normals. This way, we can better preserve the impression of sparse snow cover in regions with high variation in steepness (\cref{fig:snow} bottom).

%In contrast, the data-driven snow cover overlay is computed once for a given region in high resolution. As input, we provide the DSM tiles of the anticipated region in the highest available resolution (approximately $1 \times 1$ meters). The first compute step calculates the normals from the DSM tiles. The second compute step evaluates the albedo function depending on the interpolated normals and the user's input parameters on an even finer grid ($0.25 \times 0.25$ meters), which corresponds to the resolution of the orthophoto textures of the base map. In the compose pass, the renderer then interpolates the overlay texture rather than the normals. This way, we can better preserve the impression of sparse snow cover in regions with high variation in steepness (\cref{fig:snow} bottom).

%A limitation of the data-driven overlay is that it is impossible to compute the snow cover over a larger region because of WebGPU's texture size limitations. Given these limitations, we can currently cover a region of $2048 \times 2048$ meters ($\approx 4.2km^2$) requiring 256 MiB of GPU memory. In our implementation, we therefore combine high-detail compute snow overlays in a pre-defined focus region with lower-detail fragment shader snow in the context.

\subsection{Avalanches}
\label{sec:avalanches}

Gravitational mass flow (GMF) models compute the flow of material under gravity in a given terrain and are commonly used for snow avalanche, landslide, or rock fall simulations. Physically-based GMF models are computationally too expensive for larger-scale simulations and are also very sensitive to input parameters. Larger-scale simulations therefore usually rely on empirical models that use fewer input parameters and can approximate GMF to a certain degree. A state-of-the-art model is FlowPy~\cite{d2022flow}, which is implemented in Python and which we will use %as theoretical foundation for our implementation and 
for comparison. 

 \rev{As shown in \cref{fig:compute-pipeline}, the compute workflow first fetches elevation tiles of the DEM in the region of interest and computes the normals. By default, we operate on the recommended DEM resolution for avalanche simulations} of five to ten meters~\cite{horton2013flow}. \rev{The subsequent release point node computes, for each texel of the normal texture in parallel, whether the point is a potential release point of an avalanche.} Our release area selection is a very simple approximation, evaluating the local steepness of the terrain as only input parameter, thereby functioning very similarly as the previously described snow cover simulation. For this compute step, the user can control in which \rev{texture} resolution to evaluate potential release points and which steepness range to consider. More sophisticated models take further parameters -- like the upslope area to model the water input or geological maps describing sediment availability -- into consideration~\cite{horton2013flow}.  \rev{Alternatively, we can provide known release points encoded in a pre-computed texture, such as for historic avalanche events, as shown in \cref{fig:teaser}. The compute node outputs }a binary map indicating whether \rev{avalanche particles should} be triggered from this point or not. 
 
 \rev{The final compute node models the actual avalanche particle trajectories and the runout distance.} It takes the release \rev{point texture} and the \rev{DEM} as input, together with a variety of user-defined simulation parameters, such the ratio of persistence, which simulates some sort of momentum, and runout angle, which is a simple stopping criterion based on the angle between the current position and the release point of the \rev{particle}. \rev{The user furthermore controls the number of particles released per coded texel from the release point texture. To approximate the chaotic processes within an avalanche, we add a user-defined amount of random variation from the optimal path of a particle.}
 %
 %In the simplest setting, we trigger one \rev{particle} per release point and can thereby approximate the most likely flow of mass based on gravity from a given release point. However, avalanches accumulate a lot of mass and are therefore usually described as multi-directional flow models. To compute a multi-directional flow in parallel, the user furthermore controls how many parallel (single-directional) flows to trigger from each release point and the amount of random contribution to each flow. 
 With this Monte-Carlo approach, we can then simulate spreading mass along the flow. %\cref{fig:teaser} shows the output from a simulation run computed on DTM zoom level 15, which corresponds to input cells of approximately $5 \times 5$ meters, one release point per input cell, 256 flows per release point, 30\% random flow variation, 90\% flow persistence, and 25$^\circ$ runout angle. %Color encodes the steepness at the release point, which could be interpreted as an indicator for release risk. 
 %\rev{The avalanche trajectory compute node is performed in parallel for all released avalanche particles.}
\rev{The avalanche trajectory node computes the trajectories in parallel for all released avalanche particles.}

\begin{figure}
    \centering
    \includegraphics[alt={2D top-down view of our datasets, used for evaluating our avalanche simulation. The release points and computed avalanche path are overlaid.},width=1\linewidth]{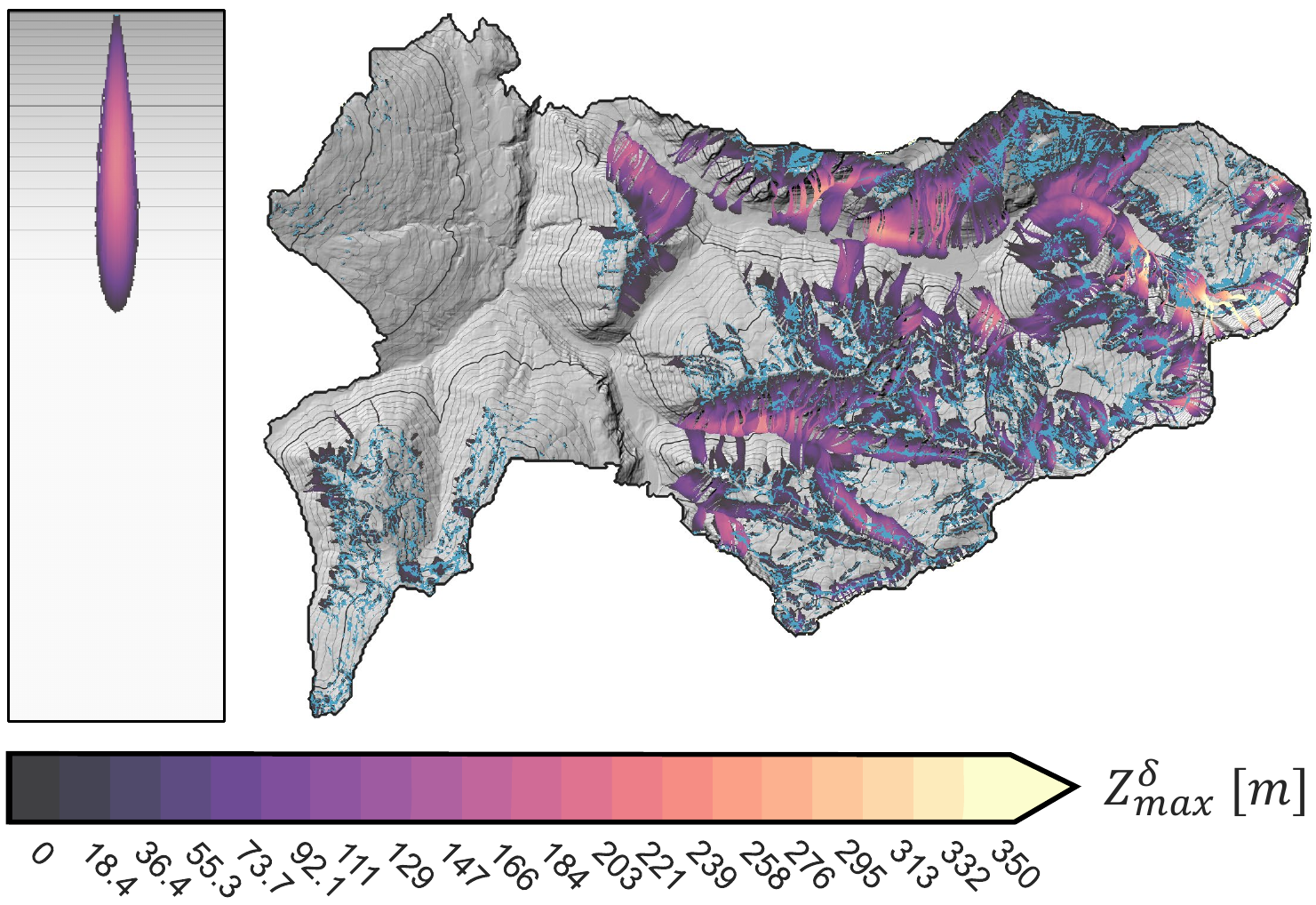}
    \caption{\emph{weBIGeo} avalanche simulation results on the \emph{parabola} (left) and \emph{vals} (right) datasets. Release points are highlighted in blue. Contour lines represent the terrain, spaced at 50m intervals. Color encodes the maximum vertical displacement per step $Z^{\delta}_{max}$, which can be interpreted as an indicator for velocity. \rev{Zoom in for more details. }}
    \label{fig:datasets}
\end{figure}

%The core compute step models the direction of the flow and decides when to stop it. It takes the release area map and the DTM as input, together with a variety of user-defined simulation parameters, such the ratio of persistence, which simulates some sort of momentum, and runout angle, which is a simple stopping criterion based on the angle between the current position and the release point of the flow. In the simplest setting, we trigger one flow per release point and can thereby approximate the most likely flow of mass based on gravity from a given release point. However, avalanches accumulate a lot of mass and are therefore usually described as multi-directional flow models. To compute a multi-directional flow in parallel, the user furthermore controls how many parallel (single-directional) flows to trigger from each release point and the amount of random contribution to each flow. With this Monte-Carlo approach, we can then simulate spreading mass along the flow. \cref{fig:teaser} shows the output from a simulation run computed on DTM zoom level 15, which corresponds to input cells of approximately $5 \times 5$ meters, one release point per input cell, 256 flows per release point, 30\% random flow variation, 90\% flow persistence, and 25$^\circ$ runout angle. Color encodes the steepness at the release point, which could be interpreted as an indicator for release risk. 

To compare the computational runtime between FlowPy's~\cite{d2022flow} Python implementation and our WebGPU compute overlays, we obtained runtime measures on two client devices: A high-end consumer \emph{desktop} (32 GB RAM, AMD Ryzen 9 3900X, NVIDIA RTX 3070 with 8 GB memory) and a commodity \emph{notebook} (16 GB RAM, Intel(R) Core(R) i7-1165G, integrated GPU with 2GB memory). For FlowPy, we used 16 CPU cores for \emph{desktop} and four cores for \emph{notebook} and fixed the parameter \emph{exp}, which controls the amount of deviation from a single-directional flow, to 8. We performed the evaluation on two data sets: \emph{Parabola} is a synthetic dataset with a parabolic slope with a resolution of $501 \times 151$ input cells, three of which are marked as release cells. \emph{Vals} is a large-scale real-world data set from Tyrol, covering an area of more than 100$km^2$. 
For our tests, simulation parameters from both data sets were read back to the CPU. We show one of these parameters in  \cref{fig:datasets}. The data sets are described in more detail in the FlowPy paper~\cite{d2022flow}. 
%For testing the performance with \emph{weBIGeo}, we systematically varied the number of flows per release point and the resolution multiplier \PK{which is what exactly?}. All performance reports were averaged from ten runs.

% choosing webigeo (model) parameters to match flowpy, then set num trajectories based on randomness
We empirically chose parameters \emph{persistence} as 90\% and \emph{randomness} as 16\% to approximate the visual results of \emph{FlowPy} using an \emph{exp} parameter of 8. Increasing \emph{randomness} introduces spread and requires computing more \rev{particles} to achieve sufficient coverage. We found that using 2048 \rev{particles} per release cell provides good visual results for the chosen randomness. All \emph{weBIGeo} performance reports were averaged from ten runs. 

A comparison between the average model runtimes shows significant speedup of \emph{weBIGeo} over \emph{FlowPy} (\cref{tab:runtimes}). On both hardware setups, \emph{weBIGeo} takes milliseconds for \emph{parabola} and seconds for \emph{vals}. In contrast, \emph{FlowPy} performs in the order of seconds for \emph{parabola} and hours for \emph{vals}. These timings exclude file read and write times. As \emph{weBIGeo} allows to visualize the simulation output directly, file export is only performed on-demand.

%For a resolution multiplier of 1 and 2048 trajectories per release point, which approximates the visual results using an \emph{exp} parameter of 8 with FlowPy, \emph{weBIGeo} required, on average, 0.0741 seconds for the entire compute pipeline on \emph{desktop} and 0.1129 seconds on \emph{notebook} for the dataset \emph{parabola}. In contrast, FlowPy takes 10.373 and 8.542 seconds on \emph{desktop} and \emph{notebook}, respectively. For the much larger data set \emph{vals}, \emph{weBIGeo} takes, on average, 3.094 and 9.344 seconds, on \emph{desktop} and \emph{notebook}, respectively. FlowPy requires more than an hour to finish on \emph{desktop} and more than four hours on \emph{notebook}. 

\begin{table}[tb]
  \caption{Model runtime comparison of \emph{FlowPy}~\cite{d2022flow} vs \emph{weBIGeo}, including speedup factor of \emph{weBIGeo} over \emph{FlowPy}.}
  \label{tab:runtimes}
  \scriptsize%
	\centering%
  \begin{tabu}{%
	r%
	r%
	r%
	r%
	r%
	}
  \toprule
   Hardware & Dataset & \emph{FlowPy} & \emph{weBIGeo} & Speedup \\
  \midrule \midrule
    \emph{desktop}  & \emph{parabola} & 10,1\,s & 13,5\,ms & 748× \\
                    & \emph{vals}     & 78,0\,min & 1,50\,s & 3120× \\
  \midrule
    \emph{notebook} & \emph{parabola} & 8,19\,s & 41,6\,ms & 197× \\
                    & \emph{vals}     & 266,7\,min & 6,74\,s & 2374× \\
 \bottomrule
 \end{tabu}%
\end{table}

\section{Discussion \& Conclusions}

Our work provides a first proof-of-concept and evaluation to assess the potential and limitations of WebGPU in the context of geographic simulation and visualization. Our first use case has shown that data-driven compute overlays can notably increase the visual quality compared to a fragment shader approach due to the independence of the overlay resolution from the base map. Our second use case shows that data-driven compute overlays have the potential to speed up gravitational mass flow simulations by orders of magnitude compared to the state-of-the-art implemented on the CPU. We have furthermore shown that client-side analysis and rendering is a feasible option even for commodity notebook hardware using Web\-GPU. The main limitation is that -- at least for overlays computed on top of high-resolution digital \rev{elevation} models as in our use cases -- the spatial extent of the overlay is limited by the GPU memory.  

\rev{While our use cases in this paper focus on interactive simulations, the principle of data-driven compute overlays can also be applied to interactive visualizations of large geographic data in the future. For instance, a compute workflow could generate heatmap overlays showing millions of user-selected points of interest or visualize the solar potential in an area of interest under different interactive scenarios.}

The choice whether to rely on client-side analysis and rendering or alternative approaches, such as pre-computed \rev{offline} tiles or server-side analysis and rendering, finally depends on multiple \rev{aspects}, such as the amount of data to be analyzed, the resources of client and server, the number of anticipated parallel users, the provided interactivity, and the bandwidth. We believe that client-side analysis and rendering is particularly promising for expert user scenarios, where we can expect users to be equipped with relatively powerful devices and a strong desire to interact intensively with simulation parameters. In the future, we therefore see WebGPU-based data-driven compute overlays as interesting option for geographic data analysis and visualization, opening the possibility for real-time inspection of simulation parameters \rev{and large geographic data} for many different scientific and public domains. 

\acknowledgments{
We thank Jan-Thomas Fischer, Felix Oesterle, \rev{Paula Spannring}, and Markus Rampp for their feedback and theoretical input for the avalanche simulations. This project was funded by NetIdee (project 6745). 
}

\balance

\bibliographystyle{abbrv-doi}

\bibliography{webigeo}
\end{document}